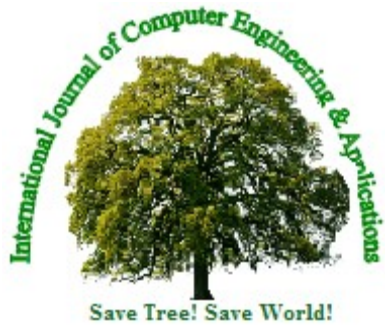

# INTEGRATED ERP SYSTEM FOR IMPROVING THE FUNCTIONAL EFFICIENCY OF THE ORGANIZATION BY CUSTOMIZED ARCHITECTURE

Dr. Suryakant B. Patil [1], Prof.Vijay S. Suryawanshi [2],
Prof. Dipali V. Suryawanshi[3], Dr. Preeti Patil[4]

[1]Professor, Head & PG Coordinator, JSPM's Imperial College of Engg. & Research, Pune, MH, India

[2]Asst. Professor, JSPM's Imperial College of Engg. & Research, Pune, MH, India

[3]Asst. Professor, JSPM's Bhivrabai Sawant Institute of Technology & Research, Pune, MH, India

[4]Dean (SA), Head & Professor, KIT College of Engineering, Kolhapur, India

**ABSTRACT:**

*An ERP is a kind of package which consist front end and backend as DBMS like a collection of DBMSs. You can create DBMS to manage one aspect of your business. For example, a publishing house has a database of books that keeps information about books such as Author Name, Title, Translator Name, etc. But this database app only helps enter books' data and search them. It doesn't help them, for example, sell books. They get or develop another DBMS database that has all the Books data plus prices, discount formulas, names of common clients, etc. Now they connect the Books database to Sales database and maybe also the inventory database. Now it's DBMS slowly turning into an ERP. They may add payroll database and connect it to this ERP. They may develop sales staff and commissions' database and connect it to this ERP and so on. In the traditional Database management system the different databases are used for the various Campuses of the JSPM Group of Education like Wagholi Campus, Tathwade Campus, Narhe Campus, Hadpsar Campuses, Bhavdhan Campus as well as Corporate office at Katraj of same organization so it is not possible to keep different databases for the same so in this paper proposed the use of Integrated Database for the Entire organization using ERP system. The Proposed ERP system applied on the existing Architecture of the JSPM Group; the marginal difference observed in the Databases need to be accessed to generate the same number of Reports when use the Traditional DBMS which end up with improvement in the Functional efficiency of Organizational Architecture.*

## [1] INTRODUCTION

Hidden ERP it is nothing but the Enterprise Resource Planning, and it is a Integrated Software for the Organizations, for example we have one Organization at Mumbai and various sub branches at all over India and also in abroad, at that time it is not possible to use various application for the various branches, so we use one Integrated Software i.e. ERP for the organization to run the Business functional operations [4,5].

ERP it is a solution designed for all the functional operations of the organization across all the locations for improving the efficiency of entire Organization [12,13].

Under ERP different modules are there as follows-
- o Finance & Accounting
- o Production Planning
- o Administration
- o HR & Payroll
- o Plant Maintenance
- o Sales & Distribution
- o Supply Chain Management
- o Customer Relationship Management
- o Business Intelligence
- o Inventory Control
- o Assets Management
- o Supply Chain Management

## [2] EXPERIMENTATION AND RESULTS

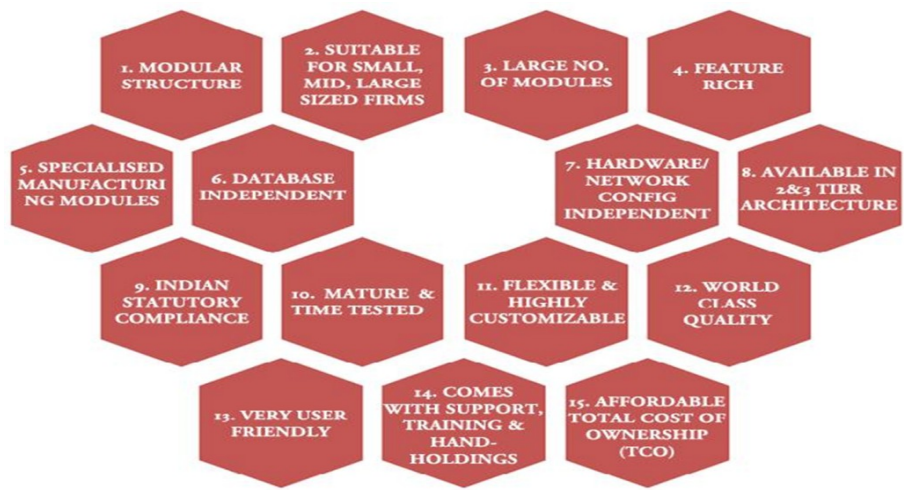

**Figure: 1. Software phases from Requirement to Post Implementation (Courtesy – PC soft ERP Solutions)**

In the market various ERP packages are there which are SAP, Oracle Apps, Baan, People Soft, When we install the ERP Package at organization require the Infrastructure [11].

Figure 1 shows various phases of ERP Life cycle from Requirement Gathering to Post Implementation phase. The Requirement Gathering phase includes Dialogue with End users; Get the proper requirements from End Users, Check with the standard information for the fulfillment of Information. Use this information for making the Data flow Diagrams depends upon the environment and Nodes select the Architecture [6, 7].

Three Categories of ERP design like –

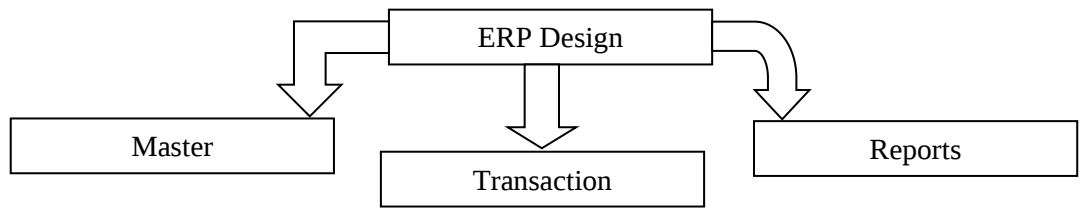

Figure: 2 Categories of ERP

Master - It is original information of the respected department or Module which are used for storing the data for doing the Routine Transaction. According to the Requirement Engineering the Phases need to be gathered as the Requirement for doing the Routine Transaction.

**Transaction –**

It is a Routine or Daily work which is depending upon the Master data; we follow this for the routine Transactions.

**Reports –**

We use the Report for generating MIS reporting or any Crystal Report and these reports are useful for the management to take a decision.

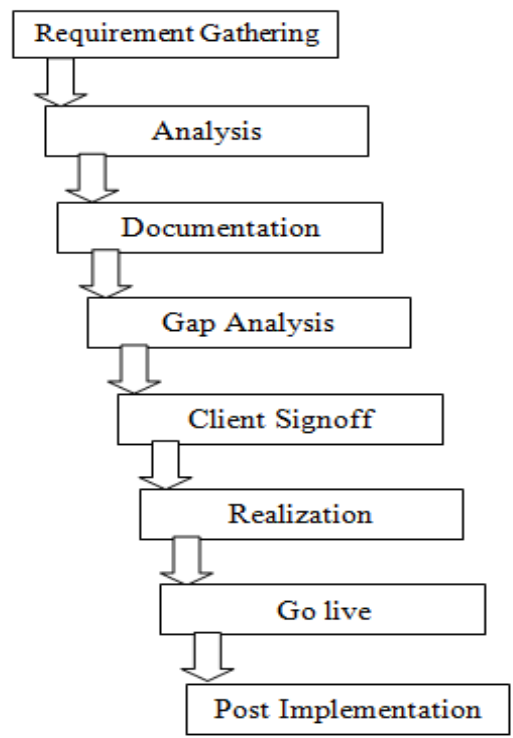

Figure: 3 ERP SDLC Phases

Requirement Gathering to Post Implementation phases are used to build any ERP application as well as Customized ERP Package as shown in figure 3.

**Phases used in the Requirement Gathering -**

Figure 4 shows the Requirement Engineering phases follow when we will work on the Requirement of any Software [8].

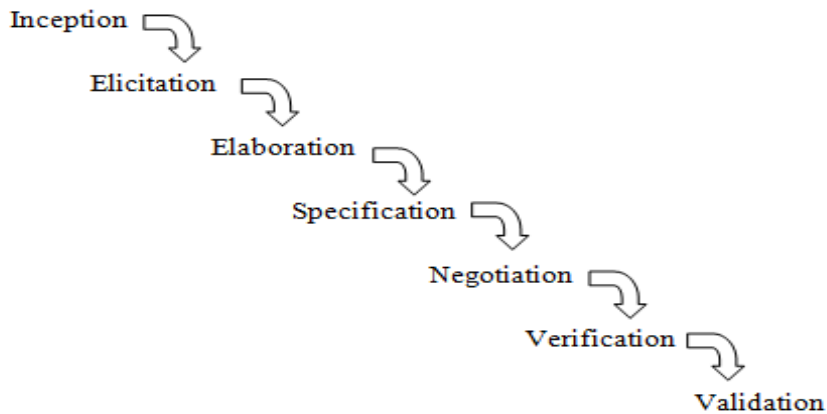

**Figure: 4 Requirement Engineering Phases**

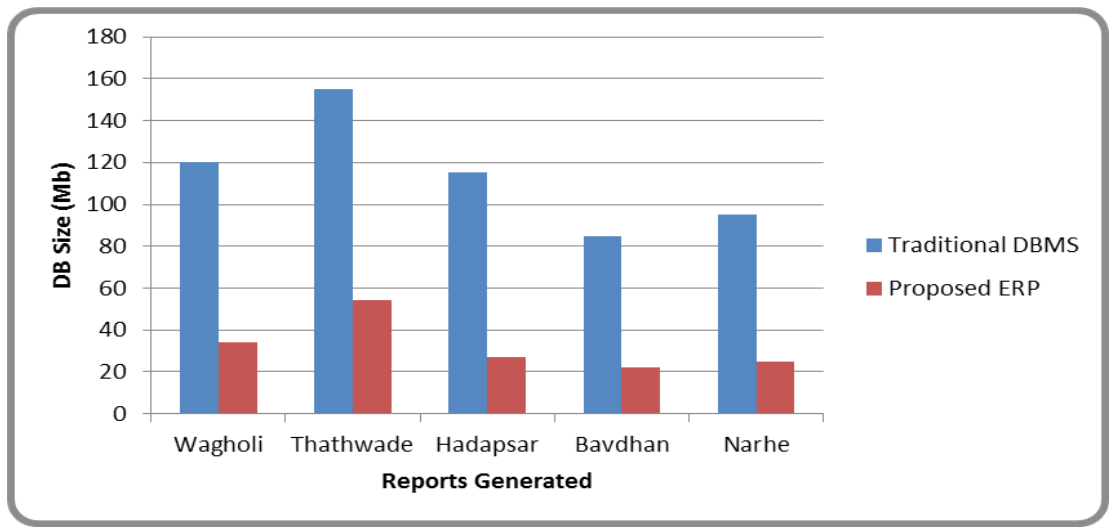

**Figure: 5 Analysis of DB Size for same No. of Report Generated**

Figure 5 focuses on the Database Size need to be accessed to create the same number of reports as a final end results in the DBMS as well as in Proposed ERP.

The fundamental advantage of ERP is that integrating myriad businesses processes saves time and expense. Management can make decisions faster and with fewer errors. Data becomes visible across the organization. Tasks that benefit from this integration include:
- Sales forecasting, this allows inventory optimization.
- Order tracking, from acceptance through fulfillment
- Revenue tracking, from invoice through cash receipt

- Matching purchase orders (what was ordered), inventory receipts (what arrived), and costing .

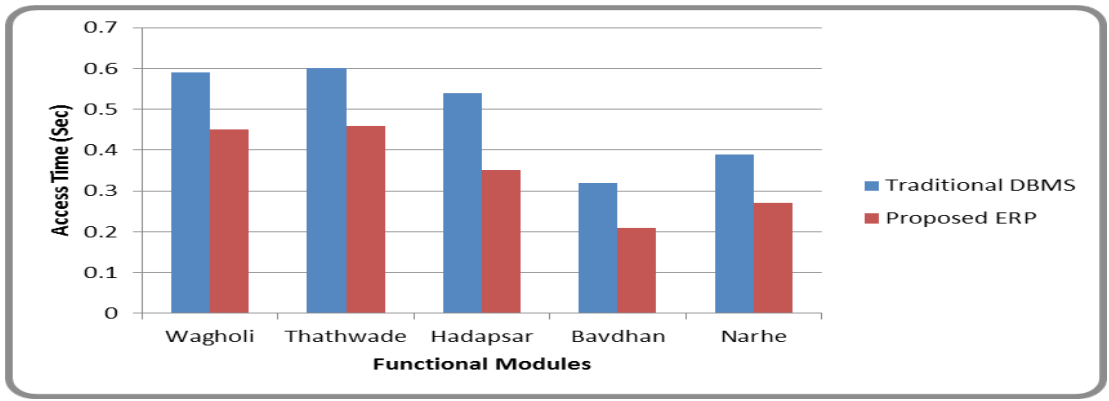

Figure: 6 Access Time Reduced in Proposed ERP

Figure 6 shows the reduced access time which need to be required to access databases by the DBMS as well as in Proposed ERP.

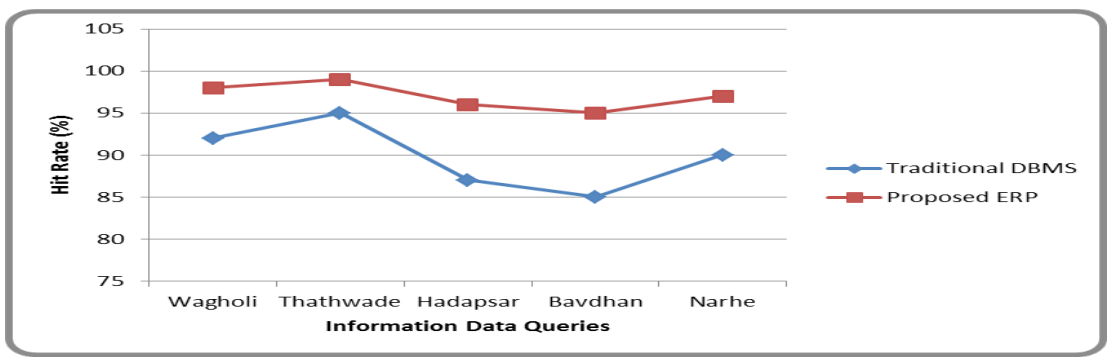

Figure: 7 Improvement in the Hit Ratio to access DB by Proposed ERP

Figure 7 shows the improvement in the Hit Ratio % to access Database the DBMS as well as in Proposed ERP.

ERP provides increased opportunities for collaboration. Data takes many forms in the modern enterprise. Documents, files, forms, audio and video, emails. Often, each data medium has its own mechanism for allowing collaboration. ERP provides a collaborative platform that lets employees spend more time collaborating on content rather than mastering the learning curve of communicating in various formats across distributed systems. ERP supports upper level management, providing critical decision making information. This decision support lets upper management make managerial choices that enhance the business.

| DBMS/ERP | Traditional DBMS | | | Proposed ERP | | |
|---|---|---|---|---|---|---|
| Functional Modules | DB Size (Mb) | Access time (sec) | Hit Rate % | DB Size (Mb) | Access time (sec) | Hit Rate % |
| **Wagholi** | 120 | 0.59 | 92 | 34 | 0.45 | 98 |
| **Thathwade** | 155 | 0.6 | 95 | 54 | 0.46 | 99 |
| **Hadapsar** | 115 | 0.54 | 87 | 27 | 0.35 | 96 |
| **Bavdhan** | 85 | 0.32 | 85 | 27 | 0.21 | 95 |
| **Narhe** | 95 | 0.39 | 90 | 25 | 0.27 | 97 |

**Table 1: Cumulative Analysis of DBMS Vs Proposed ERP**

The Cumulative Analysis of traditional DBMS Vs Proposed ERP for various functional modules at the campuses shown in Table 1.

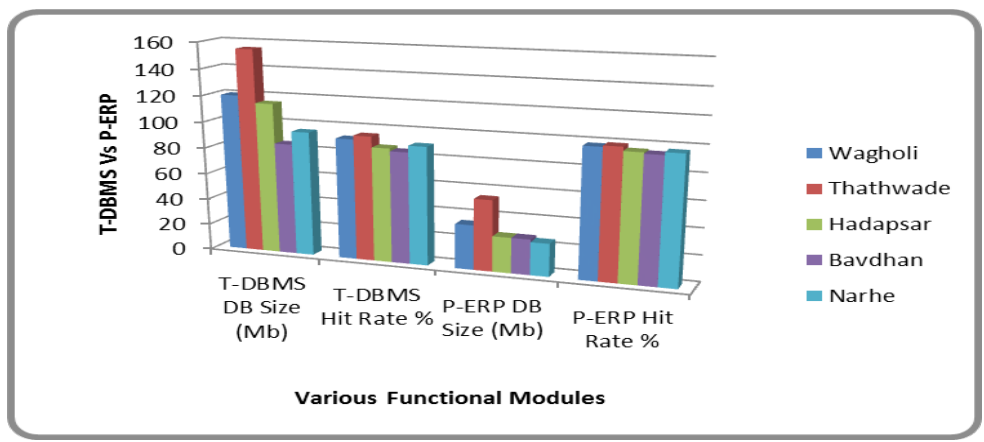

**Figure: 8 Traditional DBMS and Proposed ERP Cumulative Analysis**

Figure 8 reflects the Traditional DBMS and Proposed ERP Cumulative Analysis where traditional Database management system the different databases are used for the various Campuses of the JSPM Group of Education like Wagholi Campus, Tathwade Campus, Narhe Campus, Hadpsar Campuses, Bhavdhan Campus as well as Corporate office at Katraj of same organization so it is not possible to keep different databases for the same so in this paper proposed the use of Integrated Database for the Entire organization using ERP system.

The experimentation performed with the various set of databases and simulation wherever necessary with some real time assumptions where Proposed ERP system applied on the existing Architecture of the JSPM Group; the marginal difference observed in the Databases need to be accessed to generate the same number of Reports when use the Traditional DBMS which end up with improvement in the Functional efficiency of Organizational Efficiency.

## [3] CONCLUSION

In this paper the real time case methodology of the JSPM Group of Education refereed and the experimentation performed with the set of Databases in traditional Database Management System as well as the proposed Architecture of the ERP System. The sample Database transactions and processes on the various sub modules were observed to measure the Performance difference between the Traditional DBMS and Proposed ERP System. The Set of several queries fired between the various Campuses of the JSPM Group of Education like Wagholi Campus, Tathwade Campus, Narhe Campus, Hadpsar Campuses, Bhavdhan Campus as well as Corporate office at Katraj of same organization to access the decentralized and Centralized Databases in the Traditional DBMS and proposed ERP System respectively. With this experimentation we have proved that the required access times as well as the Centralized Database size both are reduced. Further we have proved that the Data access Hit Rate increased by 85% to 95% as the proposed ERP system enhance the Integrity along with the improved design where the Database is Divided in to the Logical Unit Numbers (LUN)

## Author[s] brief Introduction

**Dr. Suryakant B Patil**
PhD Computer Engg., M.E. CSE, B.E. CSE – 16 Years Teaching Experience
Professor & Head, JSPM's Imperial College of Engineering & Research, Pune, MH
Publications/ Papers: 71   IPR: 4 (Provisional)
Editorial Board/ Reviewer: 27

**Prof.Vijay S. Suryawanshi**
M.Tech. CSE, B.E. IT
Asst. Professor
JSPM's Imperial College of Engineering & Research, Pune, MH

**Prof. Dipali V. Suryawanshi**
M.Tech. CSE, B.E. CE
Asst. Professor
JSPM's Bhivrabai Sawant Institute of Technology & Research, Wagholi, Pune, MH, India

**Dr. Preeti Patil**
PhD Computer Engg., M.E. CSE, B.E. CSE, DCS – 12 Years Teaching Experience
Dean (SA), Professor & Head – Computer Engg. Department
KIT's College of Engineering, Kolhapur, MH
Publications/ Papers: 64   IPR: 3 (Provisional)
Editorial Board/ Reviewer: 22

## Corresponding Address-

JSPM's Imperial College of Engineering & Research, Wagholi, Pune, MH
Gat No: 720/2, Nagar Road, Pin: 412207 Mobile: 0 840 804 5555